\documentclass[prl,twocolumn,superscriptaddress,longbibliography]{revtex4-2}
\usepackage[colorlinks=true, citecolor=blue, urlcolor=blue, linkcolor=blue ]{hyperref}
\usepackage{graphicx,amsmath,amssymb,bm,amsthm,mathrsfs,orcidlink,braket}
\renewcommand{\section}[1]{{\par\it #1.---}\ignorespaces}

\begin{document}
\title{Two-atom Dicke model with atom-atom interaction}
\author{Lin Jiao}
\email{lj28@rice.edu}
\affiliation{Department of Physics and Astronomy, and Smalley-Curl Institute, Rice University, Houston, TX 77251-1892, USA}
\author{Han Pu}
\email{hpu@rice.edu}
\affiliation{Department of Physics and Astronomy, and Smalley-Curl Institute, Rice University, Houston, TX 77251-1892, USA}

\begin{abstract}
Interactions among emitters provide a powerful means of controlling collective light--matter phenomena, yet their role in superradiant criticality has not been thoroughly investigated. Here we construct a minimal model that can yield analytical insights --- a Dicke model with two interacting atoms coupled to a single mode cavity --- to study such interaction effects. We show that interaction changes the phase boundary, and may even completely suppress the atom-photon coupling threshold for superradiance and change the universality class of the phase transition. We further study the dissipative phase transition and quantum dynamics in the presence of dissipation channels such as photon loss and spin relaxation. Our results demonstrate important roles played by the atom-atom interaction and how it can be engineered to control atom-photon coupling.  
\end{abstract}
\maketitle

\textit{Introduction --- }Superradiant criticality arises from the competition between the energy cost of populating a bosonic mode and its collective coupling to matter. In the Dicke model, which describes an ensemble of two-level atoms coupled to a single mode cavity~\cite{PhysRev.93.99,HEPP1973360,PhysRevA.7.831}, this competition results in a second-order quantum phase transition in the thermodynamic limit. Later, it was found that the phase transition also occurs with a finite number of atoms (even for a single atom, in which case the Dicke model reduces to the Rabi model) in the classical oscillator limit where the atomic transition frequency far exceeds the photon frequency~\cite{PhysRevLett.90.044101,PhysRevLett.115.180404}. These studies have also been extended to open systems in the presence of various dissipation channels~\cite{PhysRevA.75.013804,PhysRevA.87.023831,PhysRevA.97.013825}. The Dicke/Rabi models and the associated superradiance phase transition have been implemented in a variety of platforms such as quantum gases inside the cavity~\cite{Baumann2010,PhysRevLett.115.230403,Zhiqiang:17,PhysRevX.8.011002,ZhangScience2021,Muniz2020,PhysRevX.11.041046,Zwettler2026}, trapped ions and superconducting Rabi simulators~\cite{PhysRevLett.121.040503,Cai2021,PhysRevX.8.021027,Braumuller2017}, NMR and circuit-QED (quantum electrodynamics) platforms~\cite{Chen2021,PhysRevLett.131.113601,PRXQuantum.5.010327}, and solid state magnetic systems~\cite{doi:10.1126/science.aat5162,MarquezPeraca2024,KimScienceAdv2025}. 

Matter interactions, which are native to many of the platforms mentioned above, introduce an independent energy scale into this picture and provide an additional control knob to light-matter coupling. Extended Dicke, spin-glass, and Dicke--Ising models predict frustration, competing magnetic and superradiant orders, and interaction-modified phase boundaries~\cite{PhysRevLett.107.277201,PhysRevLett.107.277202,PhysRevA.94.033850,PhysRevA.96.053861,PhysRevResearch.2.023131,z8gv-7yyk,qimiaowork,rao2025unilateralcriticalityphasetransition}. Experiments have explored competition between short-range interatomic and cavity-mediated interactions and demonstrated photon-mediated spin textures, exchange dynamics, correlated atom pairs, and long-range-order growth~\cite{Landig2016,Helson2023,PhysRevLett.120.223602,PhysRevLett.122.010405,PhysRevLett.132.093402,PhysRevX.15.021089}. 
However, despite these studies, a systematic and thorough investigation of the interactions' effects on superradiance criticality, as well as their interplay with dissipation, is still lacking. 

In this work, we construct a minimal model to address this question: a Dicke model with two interacting atoms in the classical oscillator limit, with and without dissipation. Despite its simplicity, which makes it analytically trackable, as we will show, this model exhibits rich physics which can be readily explored using quantum simulators based on circuit-QED~\cite{RevModPhys.93.025005,PitaVidal2024} and trapped-ion platforms~\cite{ionreview}.

\textit{Interacting two-qubit Dicke model --- }We consider two interacting two-level atoms (spins) coupled to a single mode cavity. In the main text, we will focus on the flip-flop interaction between the two spins. In the Supplemental Materials (SM)~\cite{SM}, we will discuss the effects of Ising-type interactions.
The Hamiltonian (taking $\hbar=1$) of the system reads
\begin{equation}
H=\omega a^\dagger a+\delta\Sigma_z-J(\sigma_1^+\sigma_2^-+\sigma_2^+\sigma_1^-)+g\Sigma_x(a+a^\dagger),
\label{eq:model}
\end{equation}
where $\Sigma_\mu \equiv \sigma^\mu_1+\sigma^\mu_2$, $\omega$ and $2\delta$ are the photon and bare atomic transition frequencies, respectively. Classical oscillator limit requires $\omega \ll \delta$. $J$ is the strength of the flip-flop interaction between the atoms, and $g$ the atom-photon coupling strength. For later convenience, we will introduce the dimensionless coupling strength as $\tilde{g} \equiv g/\sqrt{\omega \delta}$.

\begin{figure}[tbp]
\centering
\includegraphics[width=\linewidth]{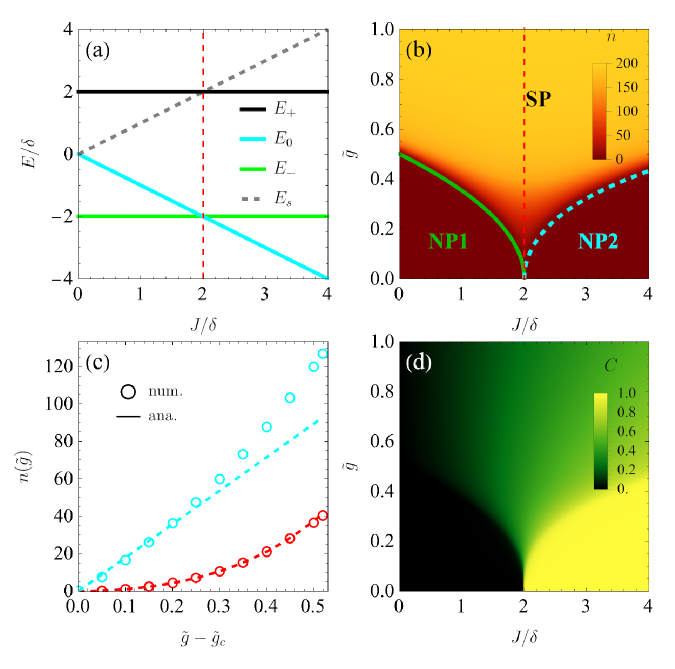}
\caption{
\textbf{(a)} Bare spin spectrum of $H_{\rm spin}$. 
\textbf{(b)} Ground-state photon number $n=\langle a^\dagger a\rangle$ in the $(J/\delta,\tilde{g})$ parameter space. The green solid and cyan dashed curves are the analytical NP--SP boundaries of Eq.~\eqref{eq:gc}.
\textbf{(c)} Photon number $n$ at $J=2\delta$ (red) and $J=3\delta$ (cyan). Circles are numerical results and dashed lines the analytical results of Eqs.~(S17) and (S21) in the SM~\cite{SM}, respectively.
\textbf{(d)} Two-spin concurrence $C$ in the same parameter plane as in \textbf{(b)}. Numerical results are obtained by exact diagonalization with $\omega=1$, $\delta=50$.}
\label{fig1}
\end{figure}

Let us first concentrate on the bare spin Hamiltonian $H_{\rm spin} = \delta\Sigma_z-J(\sigma_1^+\sigma_2^-+\sigma_2^+\sigma_1^-)$ which can be easily diagonalized. The eigenstates are simply the triplet and singlet states denoted as $|t_-\rangle=|\downarrow \downarrow\rangle$, $|t_0\rangle=(|\uparrow \downarrow\rangle+|\downarrow \uparrow \rangle)/\sqrt{2}$, $|t_+\rangle=|\uparrow \uparrow\rangle$, and $|s\rangle=(|\uparrow \downarrow\rangle-|\downarrow \uparrow \rangle)/\sqrt{2}$, with corresponding eigenenergies $E_{-}=-2\delta$, $E_{0}=-J$, $E_{+}=2\delta$, and $E_s=J$ [see Fig.~\ref{fig1}(a)]. The singlet state represents a dark state for the full Hamiltonian~(\ref{eq:model}) not coupled to the cavity as $\Sigma_x |s \rangle=0$. Consequently, we will ignore this state and from now on only consider the triplet manifold. As is obvious from Fig.~\ref{fig1}(a), the ground state of $H_{\rm spin}$ is $|t-\rangle$ for $J<2\delta$, and $|t_0 \rangle$ for $J>2\delta$. At $J=2\delta$, the ground state is doubly degenerate. As we shall see, this degeneracy significantly affects the properties of the two-atom Dicke model.

\textit{Closed-system superradiance --- }We next determine how the interaction reshapes the closed-system phase boundary. In Fig.~\ref{fig1}(b), we present the phase diagram referred to from the photon number $n=\langle a^\dag a \rangle$, obtained from Hamiltonian (\ref{eq:model}) via exact diagonalization, in the two dimensional parameter space spanned by the two dimensionless quantities $\tilde{g}$ and $J/\delta$. Depending on whether $n$ vanishes or not, we have the normal (NP) and superradiance (SP) phases. The transition between NP and SP is of second order as in conventional Dicke/Rabi models. As a function of $J/\delta$, the critical atom-photon coupling strength $\tilde{g}_c$ which delineates the phase boundary shows non-monotonic behavior. In particular, it vanishes at $J/\delta=2$, which is exactly the degenerate point of the ground state for $H_{\rm spin}$. 

By treating the atom-photon coupling as a perturbation, we can obtain the expression of $\tilde{g}_c$ analytically (see Sec. I of SM~\cite{SM} for more detail) as 
\begin{equation}
{\tilde{g}_c^2}=
\begin{cases}
({2-J/\delta})/8,& J<2\delta,\\[4pt]
{(J/\delta-4\delta/J})/{16},&J>2\delta.
\end{cases}
\label{eq:gc}
\end{equation}
At $J=2\delta$, degenerate perturbation theory shows that $\tilde{g}_c=0$, in complete agreement with the numerical results. The solid and dashed lines in Fig.~\ref{fig1}(b) represent the analytic phase boundaries.

The perturbation calculation also allows us to straightforwardly extract the critical exponents near the critical points. Here we only consider the exponent for photon number in the superradiance phase defined as 
\begin{equation}
n(\tilde{g})
\propto
({\tilde{g}-\tilde{g}_c})^\gamma,\;\;\;{\rm for}\;\;\tilde{g}>\tilde{g}_c.
\label{eq:photon_scaling}
\end{equation}
Away from the degenerate point $J \neq 2\delta$, we have $\gamma=1$, the same exponent obtained for the conventional Dicke/Rabi model; at the degenerate point $J=2\delta$, we have a different scaling with $\gamma=2$ (see SM~\cite{SM} for details). This is also confirmed by the numerical results shown in Fig.~\ref{fig1}(c). Thus, we see that the superradiance phase transition at $J=2\delta$ belongs to a different universality class compared to the conventional Dicke/Rabi model.

It is also worth noting
 that the normal phases for $J/\delta <2$ and $J/\delta >2$, labeled respectively as NP1 and NP2 in Fig.~\ref{fig1}(b), are of different nature. The former (NP1) is dominated by $|t_-,0\rangle=|t_- \rangle \otimes |0\rangle$ ($|n\rangle$ denotes the photon Fock state), while the latter (NP2) by $|t_0 ,0\rangle$. Hence in NP1 the two atoms are unentangled as manifested by the vanishing concurrence~\cite{PhysRevLett.80.2245} shown in Fig.~\ref{fig1}(d), while they are nearly maximally entangled in NP2. In the superradiance phase (SP), the atom-atom entanglement is intermediate between the two normal phases. In curious detail, in conventional superradiance systems, when crossing the phase boundary from NP to SP, the atomic excited state population increases. This is still the case crossing the boundary between NP1 and SP. However, going from NP2 to SP, the opposite happens: the excited state population decreases as shown in Fig.~S1 in the SM~\cite{SM}.

\begin{figure}[tbp]
\centering
\includegraphics[width=\linewidth]{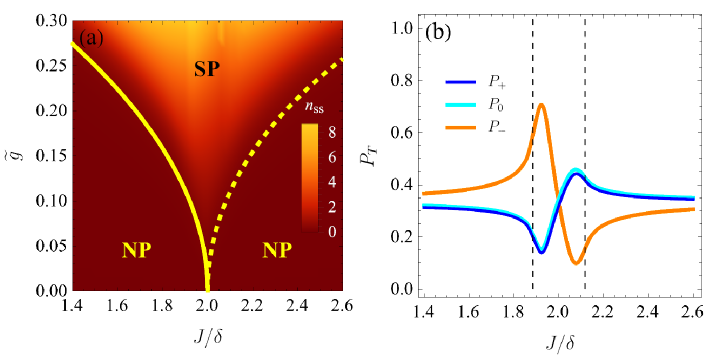}
\caption{
\textbf{(a)} Steady-state photon number $n_{\rm ss}=\langle a^\dagger a\rangle_{\rm ss}$ under cavity photon loss. The yellow solid and dashed curves denote the phase boundaries according to Eq.~\eqref{eq:gc_kappa}. 
\textbf{(b)} Steady-state triplet populations $P_\mu={\rm Tr}[(|t_\mu\rangle\langle t_\mu|\otimes I_{\rm cav})\rho_{\rm ss}]$ for $\mu=+,0,-$ along representative cut of $\tilde{g}=$0.12 are denoted by the blue, cyan, and orange curves, respectively. The vertical black dashed lines mark the phase boundaries. We set $\omega=1$, $\delta=50$, $\kappa=0.1$.}
\label{fig2}
\end{figure}

\textit{Photon loss and metastability ---}
Next, we examine the effects of dissipation. We first consider the photon loss arising from cavity decay. Under the Markovian framework, the presence of dissipation turns superradiant criticality into dissipative phase transition and Liouvillian-spectral problem~\cite{PhysRevA.75.013804,PhysRevA.87.023831,PhysRevA.97.013825,PhysRevA.98.042118}. The reduced density matrix of the system $\rho$ obeys the Master equation
\begin{equation}
\dot\rho=\mathcal L\rho=-i[H,\rho]+\kappa\mathcal D[a]\rho, \;\;
\mathcal D[o]\rho=o\rho o^\dagger-\frac{1}{2}\{o^\dagger o,\rho\},
\label{eq:pure_photon_loss_master}
\end{equation}
where $\kappa$ is the cavity decay rate.
It can be shown~\cite{SM} that, as in the conventional Dicke/Rabi models, cavity decay renormalizes the critical atom-photon coupling strength as
\begin{equation}
{\tilde{g}_{c,\kappa}^{2}}
=
\left[1+\left(\frac{\kappa}{2\omega}\right)^2\right]
{\tilde{g}_{c}^{2}},
\label{eq:gc_kappa}
\end{equation}
where $\tilde{g}_{c}$ is the corresponding closed-system threshold in Eq.~\eqref{eq:gc}. These phase boundaries are represented by the solid and dashed lines in the steady-state phase diagram in Fig.~\ref{fig2}(a).

We should emphasize that the steady state normal phases are very different from the ones for the closed system. For the closed system, NP is dominated by one of the triplet states (the ground state of $H_{\rm spin}$); in the presence of photon decay, however, NP features a significant population in all three triplet states, as shown in Fig.~\ref{fig2}(b). One can also notice that, regardless of whether the system is in NP or SP, in the steady state we have equal population in $|t_0 \rangle$ and $t_+\rangle$; at $J=2\delta$, all three triplet states are equally populated. This is quite surprising, as, at $J=2\delta$, $|t_0\rangle$ and $|t_-\rangle$ are degenerate, while $|t_+\rangle$ has higher energy when only $H_{\rm spin}$ is concerned. In the SM~\cite{SM}, we provide an explanation of all these features using a rate equation approach.

\begin{figure}[tbp]
\centering
\includegraphics[width=\linewidth]{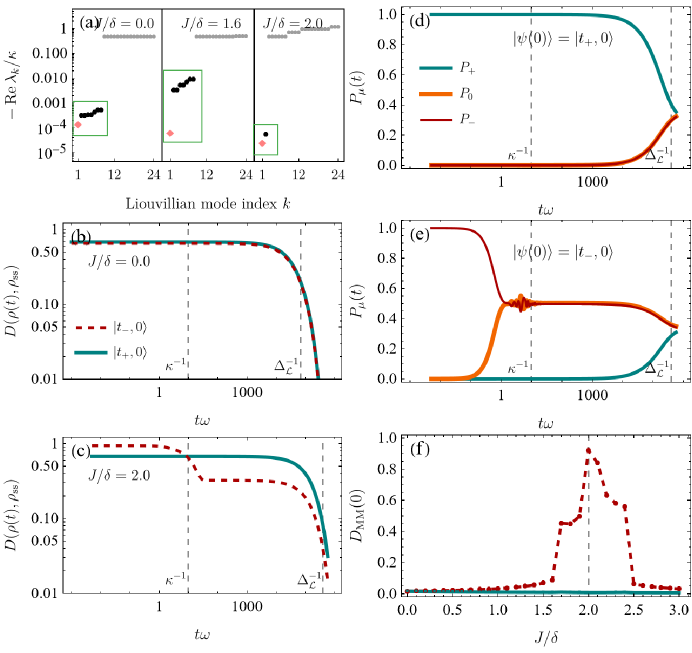}
\caption{
\textbf{(a)} Liouvillian eigenvalues $-\operatorname{Re}(\lambda_k)/\kappa$ for $J/\delta=0$, $1.6$, and $2$. The slow modes satisfying $|{\rm Re}(\lambda_k)|\ll \kappa$ and forming the metastable manifold are indicated by green boxes.
\textbf{(b)} Evolution of the normalized trace-distance $D(t)
=D_{\rm tr}[\rho(t),\rho_{\rm ss}]
/D_{\rm tr}[\rho(0),\rho_{\rm ss}]$ between the instantaneous and the steady-state density matrices for $J/\delta=0$ starting from two different initial states $|t_+,0\rangle$ (solid line) and $|t_-,0\rangle$ (dashed line). The vertical dashed lines indicate the cavity-loss time $\kappa^{-1}$ and the slow Liouvillian time $\Delta_{\mathcal L}^{-1}$.
\textbf{(c)} Same as \textbf{(b)} for $J/\delta=2$.
\textbf{(d)} Evolution of the triplet population $P_\mu(t)$ starting from the initial states $|t_+,0\rangle$ for $J/\delta=2$. \textbf{(e)} Same as \textbf{(d)} but from the initial state $|t_-,0\rangle$. 
\textbf{(f)} Initial trace distance $D_{\rm MM}(0)=\frac{1}{2}\Vert\rho(0)-\mathcal{P}_{\rm MM}\rho(0)\Vert_1$ from the metastable manifold for the initial states $|t_-,0\rangle$ (dashed line) and $|t_+,0\rangle$ (solid line), where $\mathcal{P}_{\rm MM}$ is the spectral projector onto the MM. We set $\omega=1$, $\delta=50$, $\kappa=0.1$, and $\tilde{g}=0.12$.
}
\label{fig3}
\end{figure}

The degeneracy also has an important effect on how the steady state is reached. The relaxation toward steady state is controlled by the Liouvillian spectrum $\{\lambda_k,\,k=0,1,2...\}$~\cite{PhysRevA.97.013825}, where $\lambda_i$ are eigenvalues of ${\mathcal L}$. We show representative examples of the real part of the spectrum (${\rm Re}(\lambda_k) \le 0$), which controls the relaxation time scale, in Fig.~\ref{fig3}(a). We label the eigenvalues in the descending order of ${\rm Re}(\lambda_k)$. For our system, there is one and only one null eigenvalue $\lambda_0=0$ (not shown in Fig.~\ref{fig3}(a)), indicating that there is a unique steady state (not including the singlet state which is a dark steady state we have ignored).
All spectra share a common feature: There is a cluster of eigenvalues such that $|{\rm Re}(\lambda_k)|\ll \kappa$. For reasons that will become clear later, we call the manifold spanned by this cluster the metastable manifold or MM. We define the Liouvillian gap as $\Delta_{\mathcal L} \equiv |{\rm Re}(\lambda_1)|$ which is shown as the red diamonds in Fig.~\ref{fig3}(a)). The MM is separated from the bulk spectrum for which $|{\rm Re}(\lambda_i)| \approx \kappa$.

Such separation of scales typically leads to the formation of long-lived metastable states during the relaxation toward steady state~\cite{PhysRevLett.116.240404}, since the MM indicates the existence of long-lived metastable dissipative modes. To study whether this is indeed the case, we investigate the relaxation dynamics starting from two different initial states: $|t_+,0\rangle$ and $|t_-,0\rangle$, for $\tilde{g}=0.12$ and at two different values of $J/\delta=0$ and 2. In Figs.~\ref{fig3}(b) and (c), we show how steady state is approached by plotting the trace distance
$D_{\rm tr}[\rho(t),\rho_{\rm ss}]
=\frac{1}{2}\|\rho(t)-\rho_{\rm ss}\|_1$ between the instantaneous density matrix $\rho(t)$ and the steady state density matrix $\rho_{\rm ss}$. The corresponding atomic population dynamics for $J/\delta=2$ starting from the two initial states is shown in Fig.~\ref{fig3}(d) and (e). In all four cases, only the one starting from $|t_-,0\rangle$ with $J/\delta=2$ clearly exhibits two relaxation time scales: a quasi-steady state plateau is first reached on the time scale of $\kappa^{-1}$ where populations in $|t_-\rangle$ and $|t_0\rangle$ equilibrate and that in $|t_+\rangle$ remains vanishingly small, and then the final steady state is reached on the time scale of $\Delta_{\mathcal L}^{-1}$. For the other three cases, the steady state is reached on a single time scale $\Delta_{\mathcal L}^{-1}$. 

To understand such dynamical behavior, we plot in Fig.~\ref{fig3}(f) the trace distance $D_{\rm MM}$ between the initial states ($|t_+,0\rangle$ and $|t_-,0\rangle$) and the MM at the given $J/\delta$. One can see that in the entire parameter regime, $|t_+,0\rangle$ has a very small $D_{\rm MM}$, meaning that it has a significant weight in the MM. Hence, starting from $|t_+,0\rangle$, one expects a slow relaxation on the time scale $\Delta_{\mathcal L}^{-1}$. By contrast, near the degenerate point $J/\delta=2$, $|t_-,0\rangle$ has a large $D_{\rm MM}$, indicating that it has a small weight in the MM and correspondingly a large weight in the fast modes represented by the bulk spectrum in Fig.~\ref{fig3}(a). Therefore, starting from $|t_-,0\rangle$, it will first relax to the MM on the time scale $\kappa^{-1}$ before reaching the final steady state. Physically, this is due to the following: The atom-photon coupling term couples $|t_0\rangle$ to both $|t_+\rangle$ and $|t_-\rangle$. Near $J/\delta=2$, the transition to $|t_-\rangle$ is much closer in resonance than the one to $|t_+\rangle$, making
 equilibration with $|t_+\rangle$ a very slow process. Note that the double time scale relaxation does not require exact degeneracy with $J/\delta=2$. The metastable plateau can be clearly seen for $1.5<J/\delta <2.5$, although it is most prominent at the degenerate point.

\textit{Adding local spin relaxation ---} Local spin relaxation is another common dissipation channel that is experimentally relevant for artificial or simulated spins~\cite{RevModPhys.93.025005,PhysRevA.97.013825,Barreiro2011}. Here, we investigate its effect on our model. The Master equation now takes the form
\begin{equation}
\dot\rho=-i[H,\rho]+\kappa\mathcal D[a]\rho
+\gamma_\downarrow\sum_{j=1}^{2}\mathcal D[\sigma_j^-]\rho,
\label{eq:photon_spin_loss_master}
\end{equation}
where $\gamma_\downarrow$ is the spin relaxation rate from $|\uparrow \rangle$ to $|\downarrow \rangle$. In the presence of this channel, the singlet state $|s\rangle$ is no longer an isolated dark state, as it can relax to $|t_-\rangle$ via spin relaxation. So we have to consider the full atomic Hilbert space. 

\begin{figure}[t]
\centering
\includegraphics[width=\columnwidth]{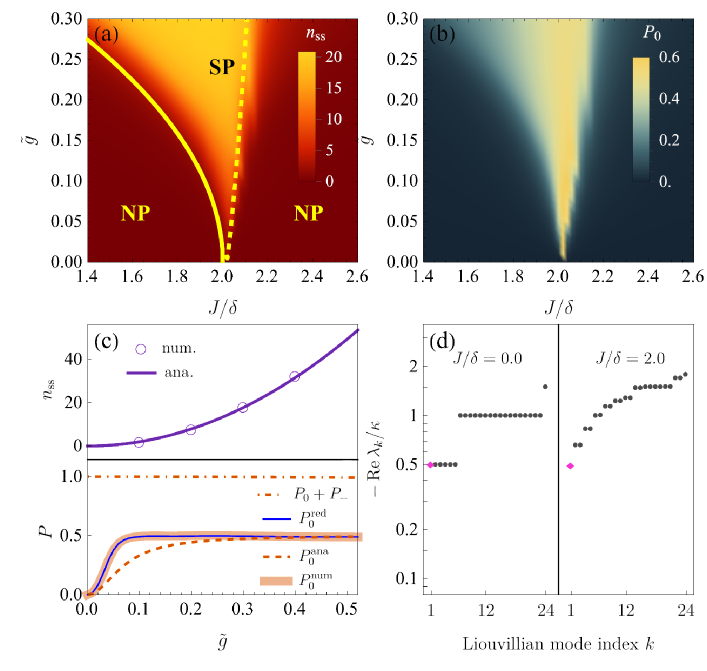}
\caption{
\textbf{(a)} Steady-state photon number $n_{\rm ss}=\langle a^\dagger a\rangle_{\rm ss}$ of the full model. The yellow lines are the analytic phase boundaries from the reduced model.
\textbf{(b)} Steady-state population $P_{0}$.
\textbf{(c)} Steady-state photon number (upper) and atomic populations (lower) at $J=2\delta$. The upper panel compares the analytical result in Eq.~\eqref{eq:spin_loss_nss} (purple solid) with the full model numerical results (open circles). The lower panel shows $P_{0}$ from the full-Liouvillian numerical (thick pale orange), reduced-Liouvillian (thin blue), and analytical (orange dashed) results; the dot-dashed curve denotes $P_{0}+P_{-}$.
\textbf{(d)} Spectrum of the full-Liouvillian $-\operatorname{Re}(\lambda_k)/\kappa$ at $\tilde g=0.12$ for $J/\delta=0$ and $2$. We set $\omega=1$, $\delta=50$, $\kappa=0.1$, $\gamma_\downarrow=0.1$ in all examples shown.}
\label{fig:spin_loss_main}
\end{figure}

Our system supports a unique steady state~\cite{SM,Evans1977,Baumgartner2008}, and the steady state phase diagram represented by the photon number is shown in Fig.~\ref{fig:spin_loss_main}(a), and the corresponding population $P_0$ for $|t_0\rangle$ is shown in Fig.~\ref{fig:spin_loss_main}(b). The NP phase is dominated by $|t_-,0\rangle$, and the SP phase has a negligible population in both $|t_+\rangle$ and $|s\rangle$. 

These results motivate us to consider a reduced Hilbert space for the atoms spanned by $\{|t_-\rangle,|t_0\rangle\}$. After projecting the atoms onto this space, the effective Hamiltonian for the reduced model reads~\cite{SM}:
\begin{equation}
H_{\rm red}=\omega a^\dagger a
+\frac{\Delta}{2}\tau_z
+\sqrt{2}g\tau_x(a+a^\dagger),
\label{eq:main_spin_loss_reduced_hamiltonian}
\end{equation}
where $\Delta \equiv 2\delta -J$, and $\tau_\mu$ are the Pauli operators acting on the basis $\{|t_-\rangle,|t_0\rangle\}$. One can see that the effective Hamiltonian (\ref{eq:main_spin_loss_reduced_hamiltonian}) takes the form of the Rabi Hamiltonian with an effective atomic transition frequency given by $\Delta$. The Master equation for the reduced model can also be straightforwardly derived as
\begin{equation}
\dot{\rho}=\mathcal L_{\rm red}\rho
=-i[H_{\rm red},\rho]
+\kappa\mathcal D[a]\rho
+\gamma_\downarrow\mathcal D[\tau_-]\rho.
\label{eq:main_spin_loss_reduced_liouvillian}
\end{equation}

Linearization of Eq.~\eqref{eq:main_spin_loss_reduced_liouvillian} around the NP state $|t_-,0\rangle$ yields the critical atom-photon coupling strength~\cite{SM}:
\[  g_c^2 = \left\{ \begin{array}{ll}  \frac{\kappa \gamma_\downarrow \left[ (\Delta-\omega)^2+(\kappa + \gamma_\downarrow)^2/4 \right] \left[ (\Delta+\omega)^2+(\kappa + \gamma_\downarrow)^2/4 \right]}{8\omega|\Delta|(\kappa + \gamma_\downarrow)^2} \,,& \Delta <0 \\
\frac{(\omega^2+\kappa^2/4)(\Delta^2+\gamma_\downarrow^2/4)}{8\omega\Delta }\,, & \Delta>0\end{array}  \right.  \]
which are plotted as solid and dashed lines in Fig.~\ref{fig:spin_loss_main}(a). 

The degenerate point $\Delta=0$ requires special treatment. From the Master equation (\ref{eq:main_spin_loss_reduced_liouvillian}), we can derive the equations of motion for the correlator $C\equiv \langle a \tau_x \rangle$ and the photon number $n=\langle a^\dag a \rangle$:
\begin{align}
 \dot C &= -(\kappa/2+\gamma_\downarrow/2+i\omega)C-i\sqrt{2}g,
 \label{eq:C-spin-degenerate}\\
 \dot n &= -\kappa n+i\sqrt{2}g(C-C^*),
 \label{eq:n-spin-degenerate}
\end{align}
which form a closed set. The steady-state solution for the photon number can be readily found as
\begin{equation}
    n_{\rm ss} = \frac{2(1+\gamma_\downarrow/\kappa)}{(\kappa + \gamma_\downarrow)^2/4+\omega^2}\,g^2.
    \label{eq:spin_loss_nss}
\end{equation}
Hence $n_{\rm ss}$ is non-vanishing as long as $g$ is non-vanishing, implying that the critical atom-photon coupling strength at the degenerate point remains at zero, just as in the previous case of a closed system or with pure photon decay. The upper panel of Fig.~\ref{fig:spin_loss_main}(c) compares $n_{\rm ss}$ numerically obtained from the full model with the analytic result of Eq.~(\ref{eq:spin_loss_nss}), and one can see that they perfectly agree with each other. 

The lower panel of Fig.~\ref{fig:spin_loss_main}(c) shows the population in $|t_0\rangle$ as a function of $\tilde{g}$ for $\Delta=0$. Three results are displayed: $P_0^{\rm num}$ from the full model, $P_0^{\rm red}$ from the reduced model, and the analytic result $P_0^{\rm ana}$ using the rate equation approach (see SM~\cite{SM} for more details). The first two are in excellent agreement, and the analytic result shows discrepancies at lower $\tilde{g}$. All these results show that $P_0$ converges to 0.5 as $\tilde{g}$ increases. In the absence of spin relaxation, all three triplet states are equally populated at the degenerate point (see Fig.~\ref{fig2}(b)). Somewhat counterintuitively, the spin relaxation can increase $P_0$, hence increasing the atom-atom entanglement. This is an interesting manifestation of the interplay between dissipation and atom-atom interaction.

Finally, in Fig.~\ref{fig:spin_loss_main}(d) we show the Liouvillian spectrum. A notable difference compared to Fig.~\ref{fig3}(a) is that here there is no clear separation of scales in ${\rm Re}(\lambda_k)$. In particular, there is no MM with $|{\rm Re}(\lambda_k)| \ll \kappa$, $\gamma_\downarrow$.

\textit{Conclusion ---}
Using a simple two-atom model, we establish matter interaction as an important factor in cavity physics. We focused on flip-flop interaction between the atoms, but other types of interaction may result in qualitatively similar physics (see Sec. IV of SM~\cite{SM} for a brief discussion of effects of Ising-type interactions). The interaction modifies the atomic spectral structure, which in turn can significantly affect the cavity response. In our model, interaction-induced degeneracy plays a particularly important role: it completely suppresses the superradiance threshold and changes the critical scaling behavior of the system. 

Dissipation is another crucial ingredient in cavity physics, determining the nature of the steady state and how the steady state is approached. In our work, we show that the interplay between matter interaction, atom-photon coupling and dissipation can lead to intriguing physics. We note that all these key ingredients can be readily realized in circuit-QED~\cite{RevModPhys.93.025005,PitaVidal2024} and trapped ion platforms~\cite{Barreiro2011,flipflop}.

Finally, to put our work in broader context, we note that cavity quantum material~\cite{10.1063/5.0083825,RevModPhys.91.025005,10.1063/5.0083825,Baydin:25} has emerged as an active field in recent years. The interaction between constituent particles is the key that gives rise to a variety of interesting properties in quantum materials and quantum many-body systems. There is still a long way to go before we have a comprehensive understanding of how cavity fields and quantum materials influence each other. The model we considered in the current work can be regarded as a toy model that will shed valuable insights towards this ultimate goal.

\textit{Acknowledgments --- }We thank Yilun Xu and Mingjian Zhu for helpful discussions.  This work is supported by the
NSF (Grant No. PHY-2513089) and the Welch Foundation (Grant No. C-1669).

\bibliography{references}

\end{document}


\title{Supplemental Material for ``Two-atom Dicke model with atom-atom interaction''}

\author{Lin Jiao}
\email{lj28@rice.edu}
\affiliation{Department of Physics and Astronomy, and Smalley-Curl Institute, Rice University, Houston, TX 77251-1892, USA}

\author{Han Pu}
\email{hpu@rice.edu}
\affiliation{Department of Physics and Astronomy, and Smalley-Curl Institute, Rice University, Houston, TX 77251-1892, USA}

\maketitle

In this Supplemental Material, we provide technical details and additional informations on the content covered in the main text. Section~I derives the closed-system critical boundary and photon-number scaling of the model. Section~II concerns our model with cavity photon loss. We derive the steady state phase boundary, establish uniqueness of the steady state within the triplet sector, and analyze the triplet populations and metastable manifold. In Section~III we construct the spin-projected reduced Liouvillian in the presence of local spin relaxation, derive the phase boundaries, and obtain the photon number and approximate spin polarization at the degenerate point $J=2\delta$. Section~IV discusses the effect of Ising-type atom--atom interaction.

\section{Closed-system criticality of the flip--flop model}
\label{sec:sm_closed}

\subsection{Triplet manifold and analytical critical boundary}
\label{sec:sm_triplet_landau}

We consider the interacting two-atom Dicke Hamiltonian introduced in the main text, setting $\hbar=1$,
\begin{equation}
H=\omega a^\dagger a+\delta\Sigma_z
-J\left(\sigma_1^+\sigma_2^-+\sigma_2^+\sigma_1^-\right)
+g\Sigma_x(a+a^\dagger),
\label{eqS:model}
\end{equation}
where $\Sigma_\mu=\sigma_1^\mu+\sigma_2^\mu$ and $\tilde g\equiv g/\sqrt{\omega\delta}$. 

The triplet--singlet basis states are denoted as $|t_-\rangle=|\downarrow \downarrow\rangle$, $|t_0\rangle=\frac{|\uparrow \downarrow\rangle+|\downarrow \uparrow\rangle}{\sqrt{2}}$, $|t_+\rangle=|\uparrow \uparrow\rangle$, $|s\rangle=\frac{|\uparrow \downarrow\rangle-|\downarrow \uparrow\rangle}{\sqrt{2}}$. We write $|t_\mu,n\rangle\equiv|t_\mu\rangle\otimes|n\rangle$, where $|n\rangle$ is a cavity Fock state. The singlet has energy $E_s=J$ and satisfies $\Sigma_x|s\rangle=0$. It is therefore a dark spectator. The cavity couples only the triplet sector. 

In the classical oscillator limit, it is justified to adopt the mean-field approach to derive the phase boundary. To this end, we replace photon operators $a$ and $a^\dag$ by a c-number $\alpha$ which can be assumed to be real. The cavity field plays the role of a transverse field $\lambda\Sigma_x$, with $\lambda\equiv 2g\alpha$. Under the basis $\{|t_-\rangle,|t_0\rangle,|t_+\rangle\}$, the corresponding Hamiltonian is
\begin{equation}
h(\lambda)=
\begin{pmatrix}
-2\delta&\sqrt{2}\lambda&0\\
\sqrt{2}\lambda&-J&\sqrt{2}\lambda\\
0&\sqrt{2}\lambda&2\delta
\end{pmatrix}.
\label{eqS:triplet_matrix}
\end{equation}
To obtain the analytical normal-to-superradiant phase boundary, we use the displaced-oscillator Landau construction appropriate under the classical oscillator limit~\cite{PhysRevLett.90.044101,PhysRevLett.115.180404}. The mean-field ground state energy is given by
\begin{equation}
E(\alpha)=\omega\alpha^2+\epsilon_0(\lambda),
\label{eqS:energy_functional}
\end{equation}
where $\epsilon_0(\lambda)$ is the lowest eigenvalue of $h(\lambda)$.

For $J<2\delta$, treating the effective transverse field as a perturbation, the unperturbed triplet ground state is $|t_-\rangle$. Nondegenerate perturbation theory gives
\begin{equation}
\epsilon_0(\lambda)=-2\delta-\frac{2\lambda^2}{2\delta-J}+O(\lambda^4),
\label{eqS:epsilon_left}
\end{equation}
and hence
\begin{equation}
E(\alpha)=-2\delta+\left(\omega-\frac{8g^2}{2\delta-J}\right)\alpha^2+O(\alpha^4).
\label{eqS:landau_left}
\end{equation}
The quadratic coefficient vanishes at
\begin{equation}
\tilde g_c^{\,2}=\frac{2-J/\delta}{8},\qquad J<2\delta.
\label{eqS:gc_left}
\end{equation}

For $J>2\delta$, the unperturbed ground state is $|t_0\rangle$, which couples to both $|t_-\rangle$ and $|t_+\rangle$. The corresponding expansion is
\begin{equation}
\epsilon_0(\lambda)=-J-\frac{2\lambda^2}{J-2\delta}-\frac{2\lambda^2}{J+2\delta}+O(\lambda^4),
\label{eqS:epsilon_right}
\end{equation}
so that
\begin{equation}
E(\alpha)=-J+\left(\omega-\frac{16Jg^2}{J^2-4\delta^2}\right)\alpha^2+O(\alpha^4).
\label{eqS:landau_right}
\end{equation}
Hence, we obtain
\begin{equation}
\tilde g_c^{\,2}=\frac{(J/\delta)^2-4}{16(J/\delta)},\qquad J>2\delta.
\label{eqS:gc_right}
\end{equation}

At $J=2\delta$, the nondegenerate expansions above are no longer valid. Projecting Eq.~\eqref{eqS:triplet_matrix} onto the degenerate subspace $\{|t_-\rangle,|t_0\rangle\}$ gives
\begin{equation}
\epsilon_0(\lambda)=-2\delta-\sqrt{2}|\lambda|+O\left(\frac{\lambda^2}{\delta}\right).
\label{eqS:epsilon_degenerate}
\end{equation}
Consequently,
\begin{equation}
E(\alpha)=-2\delta-2\sqrt{2}g|\alpha|+\omega\alpha^2+O\left(\frac{g^2\alpha^2}{\delta}\right).
\label{eqS:landau_degenerate}
\end{equation}
The linear cusp destabilizes $\alpha=0$ for any $g>0$, yielding
\begin{equation}
\tilde g_c(J=2\delta)=0.
\label{eqS:gc_degenerate}
\end{equation}
Thus, Eqs.~\eqref{eqS:gc_left} and \eqref{eqS:gc_right} reproduce Eq.~(2) of the main text on the two detuned sides of the normal-to-superradiant boundary, while the degenerate point produces a zero-threshold cavity response.


\subsection{Critical scaling and additional observables}
\label{sec:sm_closed_observables}

We now determine the photon-number scaling near the critical boundary. The three triplet eigenvalues satisfy the common secular equation
\begin{equation}
\det[\epsilon-h(\lambda)]=(\epsilon+J)(\epsilon^2-4\delta^2)-4\lambda^2\epsilon=0.
\label{eq:sm_triplet_secular_scaling}
\end{equation}
Below we denote the distance from the corresponding critical boundary by $x=\tilde g-\tilde g_c$. Our goal is to find the exponent $\gamma$ such that the photon number in the SP near the phase boundary \[ n(x) \propto x^\gamma.\]

At $J=2\delta$, a Puiseux expansion of Eq.~\eqref{eq:sm_triplet_secular_scaling} around the double root gives
\begin{equation}
\epsilon_0(\lambda)=-2\delta-\sqrt{2}|\lambda|-\frac{\lambda^2}{4\delta}+\mathcal{O}\left(\frac{|\lambda|^3}{\delta^2}\right).
\label{eq:sm_dege_spin_expansion}
\end{equation}
Substituting $\lambda=2g\alpha$ into the Landau energy gives
\begin{equation}
E(\alpha)=-2\delta-2\sqrt{2}g|\alpha|+\left(\omega-\frac{g^2}{\delta}\right)\alpha^2+\mathcal{O}\left(\frac{g^3|\alpha|^3}{\delta^2}\right).
\label{eq:sm_dege_landau_scaling}
\end{equation}
Minimizing either of the two symmetry-related branches yields
\begin{equation}
|\alpha_*|=\frac{\sqrt{2}g}{\omega-g^2/\delta}=\sqrt{\frac{2\delta}{\omega}}\frac{\tilde g}{1-\tilde g^2}.
\label{eq:sm_dege_alpha_scaling}
\end{equation}
Using $\tilde g_c=0$ at $J=2\delta$, we obtain
\begin{equation}
n_{\rm ana}(x)=|\alpha_*|^2=\frac{2\delta}{\omega}x^2\left(1+2x^2\right)+\mathcal{O}\left(\frac{\delta}{\omega}x^6\right).
\label{eq:sm_dege_n_fourth}
\end{equation}
The leading term gives $\gamma=2$. 

For $J\neq2\delta$, the normal-state root of Eq.~\eqref{eq:sm_triplet_secular_scaling} is nondegenerate and is therefore analytic in $\lambda^2$. Expanding it through fourth order and substituting $\lambda=2g\alpha$ gives
\begin{equation}
E(\alpha)=E_{\rm N}+\omega\left(1-\frac{g^2}{g_c^2}\right)\alpha^2+u_J(g)\alpha^4+\mathcal{O}(\alpha^6),
\label{eq:sm_detuned_landau}
\end{equation}
where $E_{\rm N}=-2\delta$ for $J<2\delta$, $E_{\rm N}=-J$ for $J>2\delta$, and
\begin{equation}
u_J(g)=
\begin{cases}
\dfrac{16g^4(J+2\delta)}{\delta(2\delta-J)^3}, 
&J<2\delta,\\[8pt]
\dfrac{256J(J^2+4\delta^2)g^4}{(J^2-4\delta^2)^3},
&J>2\delta.
\end{cases}
\label{eq:sm_detuned_quartic}
\end{equation}
Since $u_J(g_c)>0$, minimization on the superradiant side gives
\begin{equation}
n=\alpha_*^2=-\frac{\omega[1-g^2/g_c^2]}{2u_J(g_c)}+\mathcal{O}(x^2).
\end{equation}
Using the critical couplings obtained above, the near-critical onset becomes
\begin{equation}
n(x)=
\begin{cases}
\dfrac{4\delta(2\delta-J)}
{\omega(2\delta+J)\tilde g_c}\,x+\mathcal{O}(x^2),
&J<2\delta,\\[10pt]
\dfrac{J(J^2-4\delta^2)}
{\omega(J^2+4\delta^2)\tilde g_c}\,x+\mathcal{O}(x^2),
&J>2\delta.
\end{cases}
\label{eq:sm_detuned_linear_onset}
\end{equation}
Thus every fixed $J\neq2\delta$ exhibits the conventional linear onset with $\gamma=1$.

The main text uses the photon number to identify and the spin concurrence to characterize the normal and the superradiant phases. Here we present two additional observables that characterize the same structure from complementary viewpoints.

For a ground state $|\psi_0\rangle$, the reduced spin density matrix is
\begin{equation}
\rho_{\rm spin}=\operatorname{Tr}_{\rm cav}|\psi_0\rangle\langle\psi_0|.
\label{eqS:rho_spin}
\end{equation}
The spin--field entanglement entropy is
\begin{equation}
S=-\operatorname{Tr}(\rho_{\rm spin}\ln\rho_{\rm spin}).
\label{eqS:entropy}
\end{equation}
It is small in the low-photon regions and is enhanced when the spin and cavity sectors become hybridized. We also define the total atomic excitation population
\begin{equation}
P_e=\left\langle\psi_0\left||10\rangle\langle10|+|01\rangle\langle01|+2|11\rangle\langle11|\right|\psi_0\right\rangle .
\label{eqS:Pe}
\end{equation}
Equivalently, $P_e=P_{0}+P_s+P_+$, reduces to $P_e=P_0+P_+$ in the triplet sector.

\begin{figure}[tbp]
\centering
\includegraphics[width=\linewidth]{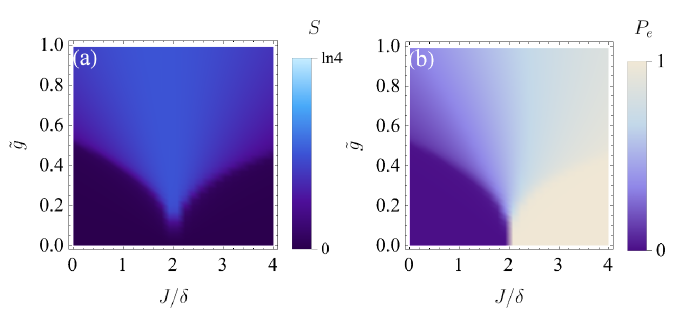}
\caption{\textbf{(a)} Spin--field entanglement entropy $S$ and \textbf{(b)} Excited atomic population $P_e$ of the closed flip--flop model. Numerical results are obtained by exact diagonalization with $\omega=1$, $\delta=50$.
}
\label{figS:additional_observables}
\end{figure}

Figure~\ref{figS:additional_observables} shows that $S$ is enhanced in the superradiant region, reflecting spin--cavity hybridization. The map of $P_e$ resolves the spin character of the two normal phases: NP1 for $J<2\delta$ is adiabatically connected to $|t_-,0\rangle$, whereas NP2 for $J>2\delta$ is connected to $|t_0,0\rangle$. The two observables therefore corroborate the interaction-driven reorganization of the normal-state manifold across $J=2\delta$.

\section{Pure photon loss}
\label{sec:sm_ph_loss}
Consider now the effect of photon loss. We will provide details on how to obtain steady state phase boundaries, proof of the uniqueness of the steady state in the triplet sector, and the atomic population distribution in the steady state.

\subsection{Dissipative superradiant boundary}

The phase boundary in Fig.~2(a) in the main text follows from the zero-frequency linear instability of the normal state within the triplet sector. Let
$P_T=\sum_{\mu=-,0,+}|t_\mu\rangle\langle t_\mu|$, $H_T=P_THP_T$, and
$\Sigma_x^{(T)}=P_T\Sigma_xP_T$. The triplet-sector master equation is
\begin{equation}
\dot{\rho}_T=-i[H_T,\rho_T]+\kappa\mathcal{D}[a]\rho_T,
\end{equation}
Here and below, $\mathcal D[O]\rho=O\rho O^\dagger-\frac12\{O^\dagger O,\rho\}$.
From this equation, the coherent cavity amplitude $\alpha=\langle a\rangle$ obeys
\begin{equation}
\dot{\alpha}=-\left(i\omega+\frac{\kappa}{2}\right)\alpha-igm_x,\qquad
m_x=\langle \Sigma_x^{(T)}\rangle.
\end{equation}
At the zero-frequency marginal mode, $\dot{\alpha}=0$, and hence
\begin{equation}
\alpha=-\frac{igm_x}{i\omega+\kappa/2}.
\end{equation}
The cavity displacement therefore generates the effective transverse field
\begin{equation}
h=g(\alpha+\alpha^\ast)=-\frac{2g^2\omega}{\omega^2+(\kappa/2)^2}m_x.
\label{eq:sm_pure_loss_cavity_feedback}
\end{equation}

We next determine the static spin response to this field. The triplet spin Hamiltonian takes the form
\begin{equation}
H_{T}(h)=H_{T}^{(0)}+h\Sigma_x^{(T)}.
\end{equation}
For a nondegenerate normal state $|\nu\rangle$, first-order perturbation theory gives
\begin{equation}
|\nu(h)\rangle=|\nu\rangle-h\sum_{\mu\neq\nu}\frac{\langle\mu|\Sigma_x^{(T)}|\nu\rangle}{E_\mu-E_\nu}|\mu\rangle+O(h^2).
\end{equation}
Consequently,
\begin{align}
m_x&=\langle\nu(h)|\Sigma_x^{(T)}|\nu(h)\rangle\nonumber\\
&=-2h\sum_{\mu\neq\nu}\frac{|\langle\mu|\Sigma_x^{(T)}|\nu\rangle|^2}{E_\mu-E_\nu}+O(h^3)\nonumber\\
&\equiv-\chi_\nu h+O(h^3).
\label{eq:sm_static_susceptibility_definition}
\end{align}
The minus sign is a convention. Thus
\begin{equation}
\chi_\nu=2\sum_{\mu\neq\nu}\frac{|\langle\mu|\Sigma_x^{(T)}|\nu\rangle|^2}{E_\mu-E_\nu}.
\label{eq:sm_static_susceptibility}
\end{equation}

For $J<2\delta$, the normal spin state is $|t_-\rangle$, which couples only to $|t_0\rangle$. Therefore
\begin{equation}
\chi_{t_-}=2\frac{|\langle t_0|\Sigma_x^{(T)}|t_-\rangle|^2}{E_0-E_-}=\frac{4}{2\delta-J}.
\label{eq:sm_susceptibility_left}
\end{equation}
For $J>2\delta$, the normal spin state is $|t_0\rangle$, which couples to both $|t_-\rangle$ and $|t_+\rangle$. Hence
\begin{align}
\chi_{t_0}&=2\left[\frac{|\langle t_-|\Sigma_x^{(T)}|t_0\rangle|^2}{E_--E_0}+\frac{|\langle t_+|\Sigma_x^{(T)}|t_0\rangle|^2}{E_+-E_0}\right]\nonumber\\
&=4\left(\frac{1}{J-2\delta}+\frac{1}{J+2\delta}\right)=\frac{8J}{J^2-4\delta^2}.
\label{eq:sm_susceptibility_right}
\end{align}

Combining $m_x=-\chi_\nu h$ with Eq.~\eqref{eq:sm_pure_loss_cavity_feedback} gives
\begin{equation}
m_x=\chi_\nu\frac{2g^2\omega}{\omega^2+(\kappa/2)^2}m_x+O(m_x^3).
\end{equation}
The normal solution $m_x=0$ becomes marginal when the linear feedback gain reaches unity,
\begin{equation}
1=\chi_\nu\frac{2g_c^2\omega}{\omega^2+(\kappa/2)^2}.
\label{eq:sm_pure_loss_instability_condition}
\end{equation}
Substituting Eqs.~\eqref{eq:sm_susceptibility_left} and \eqref{eq:sm_susceptibility_right} yields
\begin{equation}
\tilde g_{c,\kappa}^{\,2}=\left[1+\left(\frac{\kappa}{2\omega}\right)^2\right]
\begin{cases}
\displaystyle\frac{2-J/\delta}{8},&J<2\delta,\\[6pt]
\displaystyle\frac{(J/\delta)^2-4}{16(J/\delta)},&J>2\delta,
\end{cases}
\label{eq:sm_pure_loss_boundary}
\end{equation}
which is Eq.~(5) of the main text.

Equations~\eqref{eq:sm_susceptibility_left} and \eqref{eq:sm_susceptibility_right} assume nondegenerate normal states and therefore do not apply directly at $J=2\delta$. At the exact crossing, $|t_-\rangle$ and $|t_0\rangle$ are degenerate and directly coupled by $\Sigma_x^{(T)}$. Treating this degenerate triplet subspace separately shows that the normal state is unstable for any $g>0$ in the classical-oscillator limit, yielding $\tilde g_{c,\kappa}=0$.

\subsection{Triplet Liouvillian and uniqueness of the steady state}
\label{sec:sm_triplet_ness}

Here we provide details supporting the results in Fig.~2 of the main text. The two total-spin sectors remain dynamically disconnected. The triplet-sector dynamics is given by
\begin{equation}
\dot{\rho}_T=\mathcal{L}_T\rho_T=-i[H_T,\rho_T]+\kappa\mathcal{D}[a]\rho_T,
\label{eq:sm_ph_loss_LT}
\end{equation}
with
\begin{equation}
\begin{aligned}
H_T={}&\omega a^\dagger a-2\delta|t_-\rangle\langle t_-|
-J|t_0\rangle\langle t_0|+2\delta|t_+\rangle\langle t_+|\\
&+g(a+a^\dagger)\Sigma_x^{(T)} ,
\end{aligned}
\label{eq:sm_ph_loss_HT}
\end{equation}
and
\begin{equation}
\Sigma_x^{(T)}\equiv \sqrt{2}\left(|t_-\rangle\langle t_0|+|t_0\rangle\langle t_-|+|t_0\rangle\langle t_+|+|t_+\rangle\langle t_0|\right).
\label{eq:sm_ph_loss_SxT}
\end{equation} 
The tensor product with the cavity Hilbert space is implicit in the spin operators.

For $g\neq0$, the triplet-sector steady state is unique. We use the standard commutant criterion for finite-dimensional Lindblad semigroups~\cite{Evans1977,Baumgartner2008}: a sufficient condition for uniqueness is the triviality of the commutant of the Hamiltonian and the jump operators. Let $Q$ satisfy
\begin{equation}
[Q,H_T]=0,\qquad [Q,a]=0,\qquad [Q,a^\dagger]=0.
\label{eq:sm_ph_loss_commutant}
\end{equation}
The last two relations imply $Q=I_{\rm cav}\otimes Q_s$. The remaining condition requires $Q_s$ to commute with both the diagonal triplet spin Hamiltonian and the connected triplet coupling matrix $\Sigma_x^{(T)}$. Since $\Sigma_x^{(T)}$ connects the full triplet ladder $|t_-\rangle\leftrightarrow |t_0\rangle\leftrightarrow |t_+\rangle$, the only solution is $Q_s\propto I_T$. Therefore $\mathcal{L}_T$ has a unique normalised steady state, denoted by $\rho_{\rm ss}$ as in the main text. 

\subsection{Steady-state atomic population distribution}
\label{sec:sm_trip_popul}
In the regime $\delta\gg\omega,\kappa$ and $g/\delta\ll1$, the lossy cavity can be adiabatically eliminated, yielding a Pauli-rate description for the atomic populations.

Introducing $X_{\rm cav}=a+a^\dagger$, the vacuum correlation function of the damped cavity is
\begin{equation}
C_{\rm cav}(t)=\langle X_{\rm cav}(t)X_{\rm cav}(0)\rangle
=e^{-(i\omega+\kappa/2)t},\quad t>0,
\end{equation}
whose noise spectrum is
\begin{align}
S_{\rm cav}(\nu)&=2\,\operatorname{Re}\int_0^\infty dt\,e^{i\nu t}C_{\rm cav}(t)\nonumber\\
&=\frac{\kappa}{(\kappa/2)^2+(\omega-\nu)^2}.
\label{eq:sm_cavity_spectrum}
\end{align}
To second order in $g$, the cavity-induced transition rate between two spin eigenstates is
\begin{equation}
\Gamma_{i\rightarrow f}=g^2\left|\langle f|\Sigma_x|i\rangle\right|^2S_{\rm cav}(E_i-E_f).
\label{eq:rate_general}
\end{equation}
We denote the lower and upper triplet gaps by
\begin{equation}
\Delta_-\equiv E_0-E_-=2\delta-J,\quad
\Delta_+\equiv E_+-E_0=2\delta+J.
\label{eq:triplet_gaps}
\end{equation}
The four rates along the triplet ladder are
\begin{align}
\Gamma_{0\rightarrow\pm}&=\frac{2\kappa g^2}{(\Delta_\pm\pm\omega)^2+(\kappa/2)^2},\nonumber\\
\Gamma_{\pm\rightarrow0}&=\frac{2\kappa g^2}{(\Delta_\pm\mp\omega)^2+(\kappa/2)^2}.
\label{eq:triplet_rates}
\end{align}
Following the main text, we define
\begin{equation}
P_\mu\equiv\operatorname{Tr}\!\left[
\left(|t_\mu\rangle\langle t_\mu|\otimes I_{\rm cav}\right)\rho
\right],\qquad \mu=-,0,+.
\label{eq:triplet_population_definition}
\end{equation}
The corresponding Pauli rate equations are
\begin{align}
\dot P_-&=-\Gamma_{-\rightarrow0}P_-+\Gamma_{0\rightarrow-}P_0,\nonumber\\
\dot P_0&=\Gamma_{-\rightarrow0}P_--\Gamma_{0\rightarrow-}P_0-\Gamma_{0\rightarrow+}P_0+\Gamma_{+\rightarrow0}P_+,\nonumber\\
\dot P_+&=\Gamma_{0\rightarrow+}P_0-\Gamma_{+\rightarrow0}P_+,
\label{eq:triplet_rate_equations}
\end{align}
together with $P_-+P_0+P_+=1$. The stationary probability currents vanish separately along the two links,
\begin{equation}
\Gamma_{-\rightarrow0}P_-=\Gamma_{0\rightarrow-}P_0,\qquad \Gamma_{0\rightarrow+}P_0=\Gamma_{+\rightarrow0}P_+.
\end{equation}
Introducing
\begin{align}
r_-&\equiv\frac{\Gamma_{-\rightarrow0}}{\Gamma_{0\rightarrow-}}=\frac{(\Delta_--\omega)^2+(\kappa/2)^2}{(\Delta_-+\omega)^2+(\kappa/2)^2},\nonumber\\
r_+&\equiv\frac{\Gamma_{0\rightarrow+}}{\Gamma_{+\rightarrow0}}=\frac{(\Delta_+-\omega)^2+(\kappa/2)^2}{(\Delta_++\omega)^2+(\kappa/2)^2},
\label{eq:population_ratios_general}
\end{align}
one obtains $P_0=r_-P_-$ and $P_+=r_+P_0=r_-r_+P_-$. Normalization then gives the stationary populations
\begin{align}
P_-&=\frac{1}{1+r_-+r_-r_+},\\
P_0&=\frac{r_-}{1+r_-+r_-r_+},\\
P_+&=\frac{r_-r_+}{1+r_-+r_-r_+}.
\label{eq:general_triplet_populations}
\end{align}
For $J\geq0$, the upper gap $\Delta_+$ remains large compared with $\omega$ and $\kappa$. Consequently,
\begin{equation}
P_+=P_0\left[1+O\!\left(\frac{\omega}{\Delta_+}\right)\right]\simeq P_0,
\label{eq:Pt0Ptplus}
\end{equation}
throughout the parameter range shown in Fig.~2(b) of the main text.

At degeneracy $J=2\delta$, $E_-=E_0$, so that
\begin{equation}
\Gamma_{-\rightarrow0}=\Gamma_{0\rightarrow-},
\end{equation}
which implies $P_-=P_0$. Combining this relation with Eq.~\eqref{eq:Pt0Ptplus} and normalization gives, to leading order in $\omega/\delta$,
\begin{equation}
P_-\simeq P_0\simeq P_+\simeq\frac13.
\label{eq:equal13}
\end{equation}

Equation~\eqref{eq:Pt0Ptplus} explains the near coincidence of $P_0$ and $P_+$ over the range of Fig.~2(b). The near-equal population of all three triplet states is the stronger result obtained at $J=2\delta$.

\section{Photon loss and local spin relaxation}
\label{sec:sm_photon_spin_loss}
Consider now both photon loss and local spin relaxation. We will derive the reduced model, the steady state phase boundary, and the atomic population distribution at the degenerate point.

\subsection{Spin-projected reduced Liouvillian}

The Master equation takes the form
\begin{equation}
\dot{\rho}=-i[H,\rho]+\kappa\mathcal D[a]\rho+\gamma_\downarrow\sum_{j=1}^{2}\mathcal D[\sigma_j^-]\rho.
\label{eq:sm_photon_spin_loss_master}
\end{equation}
Unlike photon loss, the local spin jump operators do not conserve total spin and therefore connect the triplet and singlet sectors. The singlet-vacuum is no longer an independent dark state, and the full Liouvillian selects a unique steady state. Numerically, this steady state is predominantly supported in the spin subspace $\mathrm{span}\{|t_-\rangle,|t_0\rangle\}$, which motivates the construction of a reduced model as follows.

We define the spin and spin--cavity projectors
\begin{equation}
\mathcal P_{\rm red}^{\rm spin}=|t_-\rangle\langle t_-|+|t_0\rangle\langle t_0|,\qquad
\mathcal P_{\rm red}=\mathcal P_{\rm red}^{\rm spin}\otimes I_{\rm cav}.
\label{eq:sm_spin_loss_projector}
\end{equation}
Only the spin sector is projected, while the full cavity Fock space is retained. The reduced Hamiltonian is $H_{\rm red}=\mathcal P_{\rm red}H\mathcal P_{\rm red}$. With $\Delta\equiv2\delta-J$, using $E_-=-2\delta$, $E_0=-J$, and
\begin{equation}
\mathcal P_{\rm red}^{\rm spin}\Sigma_x\mathcal P_{\rm red}^{\rm spin}
=\sqrt{2}\left(|t_-\rangle\langle t_0|+|t_0\rangle\langle t_-|\right),
\label{eq:sm_spin_loss_sigma_x_projection}
\end{equation}
one obtains, after subtracting the constant energy $-2\delta$,
\begin{equation}
\begin{aligned}
H_{\rm red}={}&\omega a^\dagger a+\Delta|t_0\rangle\langle t_0|\\
&+\sqrt{2}g(a+a^\dagger)
\left(|t_-\rangle\langle t_0|+|t_0\rangle\langle t_-|\right).
\end{aligned}
\end{equation}
Defining Pauli operators $\tau_z=|t_0\rangle\langle t_0|-|t_-\rangle\langle t_-|$, $\tau_x=|t_-\rangle\langle t_0|+|t_0\rangle\langle t_-|$, $\tau_-=|t_-\rangle\langle t_0|$, $\tau_+=|t_0\rangle\langle t_-|$, the reduced Hamiltonian can be written as
\begin{equation}
H_{\rm red}=\omega a^\dagger a+\frac{\Delta}{2}\tau_z+\sqrt{2}g\tau_x(a+a^\dagger),
\label{eq:sm_spin_loss_rabi_form}
\end{equation}
up to an irrelevant constant. Thus the flip--flop interaction tunes the detuning of an effective two-level Rabi model formed by the collective spin states $|t_-\rangle$ and $|t_0\rangle$, with degeneracy at $J=2\delta$.

We next project the dissipators. The cavity-loss jump operator is unchanged because it acts only on the cavity Hilbert space. The local spin-relaxation operators satisfy
\begin{equation}
\mathcal P_{\rm red}^{\rm spin}\sigma_1^-\mathcal P_{\rm red}^{\rm spin}
=\mathcal P_{\rm red}^{\rm spin}\sigma_2^-\mathcal P_{\rm red}^{\rm spin}
=\frac{1}{\sqrt{2}}\tau_-.
\label{eq:sm_spin_loss_sigma_minus_projection}
\end{equation}
Therefore the two microscopic decay channels combine as
\begin{equation}
\gamma_\downarrow\sum_{j=1}^{2}\mathcal D\!\left[
\mathcal P_{\rm red}^{\rm spin}\sigma_j^-\mathcal P_{\rm red}^{\rm spin}
\right]\rho=\gamma_\downarrow\mathcal D[\tau_-]\rho .
\label{eq:sm_spin_loss_effective_decay}
\end{equation}
The final reduced Liouvillian is
\begin{equation}
\mathcal L_{\rm red}\rho=-i[H_{\rm red},\rho]+\kappa\mathcal D[a]\rho+\gamma_\downarrow\mathcal D[\tau_-]\rho.
\label{eq:sm_spin_loss_reduced_liouvillian}
\end{equation}


\subsection{Linear stability of the reduced model}

Let $G=\sqrt{2}g$, $K=\kappa/2$, and $R=\gamma_\downarrow/2$. Using the cavity quadratures $x=\alpha+\alpha^\ast$ and $p=-i(\alpha-\alpha^\ast)$ and defining $s_\mu=\langle\tau_\mu\rangle$, we linearize Eq.~\eqref{eq:sm_spin_loss_reduced_liouvillian} about the normal fixed point $|t_-,0\rangle$:
\begin{align}
\dot x&=-Kx+\omega p,&\dot p&=-\omega x-Kp-2Gs_x,\nonumber\\
\dot s_x&=-Rs_x-\Delta s_y,&\dot s_y&=2Gx+\Delta s_x-Rs_y.
\label{eq:sm_linearized_eom}
\end{align}
Writing $\mathbf v=(x,p,s_x,s_y)^T$, these equations take the form $\dot{\mathbf v}=\mathbf M\mathbf v$, with
\begin{equation}
\mathbf M=\begin{pmatrix}-K&\omega&0&0\\-\omega&-K&-2G&0\\0&0&-R&-\Delta\\2G&0&\Delta&-R\end{pmatrix}.
\end{equation}
For a perturbation $\mathbf v(t)\propto e^{\lambda t}$, the characteristic equation is
\begin{equation}
\left[(\lambda+K)^2+\omega^2\right]\left[(\lambda+R)^2+\Delta^2\right]-4\Delta\omega G^2=0.
\label{eq:sm_characteristic_equation}
\end{equation}
The normal fixed point is linearly stable when all eigenvalues satisfy $\operatorname{Re}\lambda<0$. At the stability boundary, an eigenvalue reaches the imaginary axis. Setting $\lambda=i\nu$ and separating Eq.~\eqref{eq:sm_characteristic_equation} into real and imaginary parts gives
\begin{align}
&\left(K^2+\omega^2-\nu^2\right)\left(R^2+\Delta^2-\nu^2\right)-4KR\nu^2-4\Delta\omega G^2=0,\label{eq:sm_stability_real}\\
&2\nu\left[R\left(K^2+\omega^2-\nu^2\right)+K\left(R^2+\Delta^2-\nu^2\right)\right]=0.\label{eq:sm_stability_imag}
\end{align}
Eq.~(\ref{eq:sm_stability_imag}) yields the following solutions:
\begin{equation}
    \nu_1=0\,,\;\;\;\nu_2^2=\frac{R(K^2+\omega^2)+K(R^2+\Delta^2)}{R+K}
\end{equation}

\begin{itemize}
    \item 

For $\Delta>0$ ($J<2\delta$), we need to take $\nu_1=0$ as the physical marginal solution (solution $\nu_2$ leads to unphysical situation with $g^2<0$). Equation~\eqref{eq:sm_stability_real} then yields
\begin{equation}
g_{c}^{2}=\frac{(\omega^2+\kappa^2/4)(\Delta^2+\gamma_\downarrow^2/4)}{8\Delta\omega}.
\label{eq:sm_gc_left}
\end{equation}
Thus, the normal fixed point in this case loses stability through a zero-frequency mode, with a real eigenvalue reaching $\lambda=0$. In the lossless limit $K,R\rightarrow0$, Eq.~\eqref{eq:sm_gc_left} reduces to the closed-system result $g_{c}^2=\omega\Delta/8$.

\item For $\Delta<0$ ($J>2\delta$), we should take $\nu_2$ as $\nu_1=0$ leads to the unphysical $g^2<0$. 
Defining $\Xi=K^2+\omega^2-R^2-\Delta^2$, one finds
\begin{equation}
\begin{aligned}
K^2+\omega^2-\nu_{2}^2&=\frac{K\Xi}{K+R},\\
R^2+\Delta^2-\nu_{2}^2&=-\frac{R\Xi}{K+R}.
\end{aligned}
\end{equation}
Substitution into Eq.~\eqref{eq:sm_stability_real}, together with
\begin{align}
\Xi^2+4(K+R)^2\nu_{2}^2&=\nonumber\\
\left[(\Delta-\omega)^2+(K+R)^2\right]&\left[(\Delta+\omega)^2+(K+R)^2\right],
\end{align}
gives
\begin{equation}
\begin{aligned}
g_{c}^{2}={}&
\frac{\kappa\gamma_\downarrow}
{8|\Delta|\omega(\kappa+\gamma_\downarrow)^2}\\
&\times\left[(\Delta-\omega)^2
+\frac{(\kappa+\gamma_\downarrow)^2}{4}\right]\\
&\times\left[(\Delta+\omega)^2
+\frac{(\kappa+\gamma_\downarrow)^2}{4}\right].
\end{aligned}
\label{eq:sm_gc_right}
\end{equation}

\end{itemize}



\subsection{Spin polarization near the interaction-tuned crossing}

We now consider the steady-state atomic population distribution near $J=2\delta$, following a similar rate equation approach discussed above. Eqs.~\eqref{eq:sm_cavity_spectrum} and \eqref{eq:rate_general} give
\begin{align}
\Gamma_{-\rightarrow0}(\Delta)&=G^2S_{\rm cav}(-\Delta)=\frac{2G^2K}{K^2+(\omega+\Delta)^2},\label{eq:sm_tauz_rate_forward}\\
\Gamma_{0\rightarrow-}(\Delta)&=G^2S_{\rm cav}(\Delta)=\frac{2G^2K}{K^2+(\omega-\Delta)^2}.\label{eq:sm_tauz_rate_backward}
\end{align}
Including the spin relaxation $|t_0\rangle\rightarrow|t_-\rangle$, the population equation becomes
\begin{equation}
\dot P_0=\Gamma_{-\rightarrow0}(\Delta)P_--\left[\Gamma_{0\rightarrow-}(\Delta)+\gamma_\downarrow\right]P_0.
\label{eq:sm_tauz_population_detuned}
\end{equation}
Using $P_0+P_-=1$, the stationary populations are
\begin{align}
P_0^{\rm ss}(\Delta)&=\frac{\Gamma_{-\rightarrow0}(\Delta)}{\Gamma_{-\rightarrow0}(\Delta)+\Gamma_{0\rightarrow-}(\Delta)+\gamma_\downarrow},\label{eq:sm_tauz_P0_detuned}\\
P_-^{\rm ss}(\Delta)&=\frac{\Gamma_{0\rightarrow-}(\Delta)+\gamma_\downarrow}{\Gamma_{-\rightarrow0}(\Delta)+\Gamma_{0\rightarrow-}(\Delta)+\gamma_\downarrow}.\label{eq:sm_tauz_Pminus_detuned}
\end{align}

This result also explains the asymmetric population profile around the crossing. The rate $\Gamma_{-\rightarrow0}$ is resonantly enhanced near $\Delta=-\omega$, corresponding to $J\simeq2\delta+\omega$, and therefore increases $P_0$. By contrast, $\Gamma_{0\rightarrow-}$ is enhanced near $\Delta=+\omega$ and depletes $P_0$. The maximum of $P_0$ can consequently lie slightly on the $J>2\delta$ side.

At $\Delta=0$, the two rates become equal:
\begin{equation}
\begin{aligned}
\Gamma_{-\rightarrow0}(0)&=\Gamma_{0\rightarrow-}(0)=\Gamma_{\rm cav},\\
\Gamma_{\rm cav}&=G^2S_{\rm cav}(0)=\frac{2G^2K}{K^2+\omega^2}.
\end{aligned}
\label{eq:sm_tauz_rate}
\end{equation}
Equations~\eqref{eq:sm_tauz_P0_detuned} and \eqref{eq:sm_tauz_Pminus_detuned} then reduce to
\begin{equation}
P_0^{\rm ss}=\frac{\Gamma_{\rm cav}}{2\Gamma_{\rm cav}+\gamma_\downarrow},\qquad P_-^{\rm ss}=\frac{\Gamma_{\rm cav}+\gamma_\downarrow}{2\Gamma_{\rm cav}+\gamma_\downarrow}.
\label{eq:sm_tauz_populations}
\end{equation}
Local spin relaxation therefore biases the steady state toward $|t_-\rangle$. When $\Gamma_{\rm cav}\gg\gamma_\downarrow$,
\begin{equation}
P_-^{\rm ss}\simeq P_0^{\rm ss}\simeq\frac{1}{2}.
\label{eq:sm_tauz_saturation}
\end{equation}
Relative to the pure-photon-loss triplet steady state, local relaxation removes $|t_+\rangle$ from the dominant steady-state subspace while the singlet population remains negligible. The resulting redistribution allows $P_0$ to increase from approximately $1/3$ toward $1/2$, as shown in Fig.~4(c) of the main text.

\section{Ising-type interactions}
\label{sec:sm_anisotropic}

\begin{figure*}[tbp]
\centering
\includegraphics[width=0.98\linewidth]{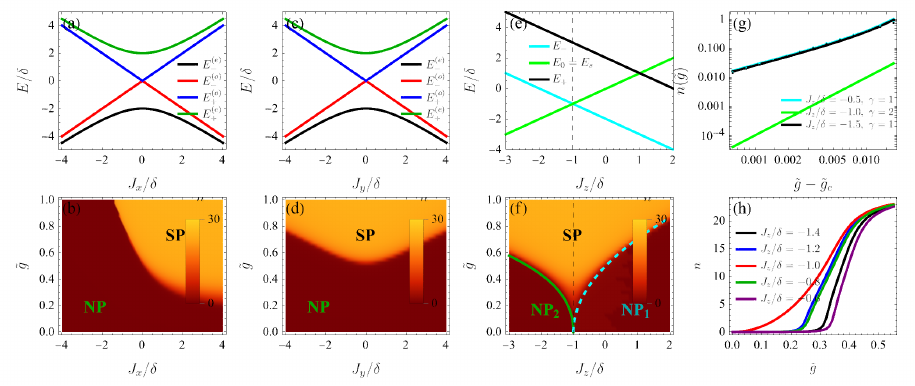}
\caption{Spin spectrum at $g=0$ and ground-state photon number $n$ for the \textbf{(a,b)} $XX$, \textbf{(c,d)} $YY$ and \textbf{(e,f)} $ZZ$ coupling. In \textbf{(a)} and \textbf{(c)}, $e$ and $o$ denote the even sector $\{|00\rangle,|11\rangle\}$ and odd sector $\{|10\rangle,|01\rangle\}$, respectively. 
\textbf{(f)} The green solid and cyan dashed curves denote the two analytical phase boundaries in Eqs.~\eqref{eqS:ZZ_gc_left} and \eqref{eqS:ZZ_gc_right}, and the vertical dashed line marks $J_z/\delta=-1$. 
\textbf{(g)} Critical onset of the photon population near the $ZZ$ crossing. 
\textbf{(h)} Representative photon-number cuts across the $ZZ$ critical region. All numerical results use $\omega=1$, $\delta=50$.}
\label{figS:anisotropic}
\end{figure*}

We now examine other types of atom-atom interactions. Specifically, consider the Ising-type interaction with the corresponding Hamiltonian
\begin{equation}
H_\mu=\omega a^\dagger a+\delta\Sigma_z-J_\mu\sigma_1^\mu\sigma_2^\mu+g\Sigma_x(a+a^\dagger),\qquad \mu=x,y,z .
\label{eqS:anisotropic_model}
\end{equation}
The cavity coupling remains the same as $\Sigma_x(a+a^\dagger)$ which will be treated as a perturbation. For a nondegenerate spin ground state $|0\rangle$ satisfying $\langle0|\Sigma_x|0\rangle=0$, expanding the ground-state energy as $E(\alpha)=E_0+c_2\alpha^2+\mathcal{O}(\alpha^4)$ gives
\begin{equation}
c_2=\omega-4g^2\chi_x,\qquad 
\chi_x=\sum_{m\neq0}\frac{|\langle m|\Sigma_x|0\rangle|^2}{E_m-E_0}.
\label{eqS:susceptibility}
\end{equation}

For the $XX$ and $YY$ interactions (i.e., for $\mu=x$ or $y$ in Eq.~(\ref{eqS:anisotropic_model})), the spin Hamiltonians are isospectral. Their eigenenergies are
\begin{equation}
\frac{E_\pm^{(e)}}{\delta}=\pm\sqrt{4+(J_\mu/\delta)^2},\quad 
\frac{E_\pm^{(o)}}{\delta}=\pm |J_\mu/\delta|,\quad \mu=x,y,
\label{eqS:XXYY_spectrum}
\end{equation}
where the even sector is spanned by $\{|\downarrow \downarrow\rangle,|\uparrow \uparrow \rangle\}$ and the odd sector is spanned by $\{|\uparrow \downarrow \rangle,|\downarrow \uparrow \rangle\}$. The labels in Figs.~\ref{figS:anisotropic}(a) and \ref{figS:anisotropic}(c) denote the sorted lower and upper branches within each sector.

Although the $XX$ and $YY$ spin Hamiltonians are isospectral, the superradiant instability is not determined by eigenvalues alone. It also depends on how the spin eigenstates are coupled by the fixed cavity operator $\Sigma_x$. The relevant projected-coupling weight is
\begin{equation}
M_m^{(\mu)}=|\langle m_\mu|\Sigma_x|0_\mu\rangle|^2,
\label{eqS:active_weights}
\end{equation}
where $|0_\mu\rangle$ and $|m_\mu\rangle$ are eigenstates of the corresponding spin Hamiltonian. Thus two models with identical spectra can have different phase diagrams if their eigenstates have different matrix elements under $\Sigma_x$. This explains why the phase diagrams (represented by photon number) in Figs.~\ref{figS:anisotropic}(b) and \ref{figS:anisotropic}(d) are not identical.

The $ZZ$ interaction provides a particularly transparent example because it reorders diagonal product-state levels without hybridizing them. The spin Hamiltonian is
\begin{equation}
H_{\rm spin}^{ZZ}=\delta\Sigma_z-J_z\sigma_1^z\sigma_2^z.
\label{eqS:Hzz}
\end{equation}
Its eigenstates are the triplet and singlet states, with the corresponding eigenenergies $E_{-}=-2\delta-J_z$, $E_{+}=2\delta-J_z$, and
$E_{0}=E_s=J_z$. The degeneracy between $|t_0 \rangle$ and $|s\rangle$ is not interesting in our case, since $|s\rangle$ is a dark state not coupled to the cavity and can be neglected. As shown in Fig.~\ref{figS:anisotropic}(e), $E_-$ crosses $E_0$ at $J_z/\delta=-1$. This degeneracy will have a significant effect on the system just as the degenerate point ($J=2\delta$) in the case of flip-flop interaction. 

For $J_z/\delta>-1$, the normal state is $|t_-,0\rangle$. The leading cavity-coupled channel is $|t_-\rangle\rightarrow|t_0\rangle$, and the second-order perturbation theory gives
\begin{equation}
\tilde{g}_c^2=\frac{1+J_z/\delta}{4}.
\label{eqS:ZZ_gc_right}
\end{equation}
For $J_z/\delta<-1$, $|t_0\rangle$ couples to both $|t_-\rangle$ and $|t_+\rangle$, contributing to the instability and yielding
\begin{equation}
\tilde{g}_c^2
=-\frac{(J_z/\delta)^2-1}{8(J_z/\delta)}.
\label{eqS:ZZ_gc_left}
\end{equation}
The two analytical branches in Eqs.~\eqref{eqS:ZZ_gc_right} and \eqref{eqS:ZZ_gc_left} vanish at $J_z/\delta=-1$, producing the cusp in the $ZZ$ phase boundary shown in Fig.~\ref{figS:anisotropic}(f). The normal regions labeled $\mathrm{NP}_1$ and $\mathrm{NP}_2$ correspond, respectively, to the $|t_0,0\rangle$ for $J_z/\delta<-1$ and the $|t_-,0\rangle$ for $J_z/\delta>-1$.

The photon-number scaling near the $ZZ$ crossing follows the same form as Eq.~(3) in the main text. Figures~\ref{figS:anisotropic}(g) and \ref{figS:anisotropic}(h) show that $\gamma\simeq2$ at $J_z/\delta=-1$, while generic points away from it give $\gamma\simeq1$. The same change of onset exponent, therefore, arises for microscopically different interactions whenever two low-energy state are tuned into degeneracy while retaining a nonzero matrix element under the cavity-coupling operator.

\bibliography{references}